\documentclass[a4paper,USenglish]{article}

\usepackage{amsmath,amssymb, amsthm, mathtools}

\usepackage[utf8]{inputenc}

\usepackage{textcomp,marvosym}

\usepackage{cite}

\usepackage{nameref,hyperref}

\usepackage{microtype}
\DisableLigatures[f]{encoding = *, family = * }

\usepackage[table]{xcolor}

\usepackage{array}

\usepackage{xspace}

\usepackage[ruled, vlined, linesnumbered]{algorithm2e}

\usepackage{color}
\usepackage{subcaption}
\usepackage{afterpage}
\usepackage{authblk}

\usepackage{numprint}
\npdecimalsign{.}
\nprounddigits{0}

\date{}

\usepackage{lastpage,fancyhdr,graphicx}
\usepackage{epstopdf}

\usepackage{wrapfig}
\usepackage{placeins}

\newcommand{\bigO}[1]{\mathcal{O}(#1)}

\newcommand{\angstrom}{\text{\normalfont\AA}}

\newcommand{\iec}{\textit{i.\,e., }}
\newcommand{\ie}{\textit{i.\,e.}\xspace}
\newcommand{\egc}{\textit{e.\,g., }}
\newcommand{\eg}{\textit{e.\,g.}\xspace}
\newcommand{\etal}{\textit{et al.}\xspace}
\newcommand{\wrt}{w.\ r.\ t.\ }
\newcommand{\NWK}{\textsf{MOBi}\xspace}
\newcommand{\DGSOL}{\textsf{DGSOL}\xspace}
\newcommand{\DISCO}{\textsf{DISCO}\xspace}
\newcommand{\rmsd}{RMSD\xspace}

\newtheorem{definition}{Definition}

\title{Maxent-stress Optimization of 3D Biomolecular Models} 

\author[1]{Michael Wegner}
\author[2]{Oskar Taubert}
\author[3]{Alexander Schug}
\author[1]{Henning Meyerhenke}
\affil[1]{Institute of Theoretical Informatics, Karlsruhe Institute of Technology (KIT), Karlsruhe, Germany}
\affil[2]{Department of Physics, Karlsruhe Institute of Technology (KIT), Karlsruhe, Germany}
\affil[3]{Steinbuch Centre for Computing, Karlsruhe Institute of Technology (KIT), Karlsruhe, Germany}

\begin{document}
\maketitle 

\begin{abstract}
Knowing a biomolecule's structure is inherently linked to and a prerequisite for any detailed understanding of its function. Significant effort has gone into developing technologies for structural characterization. These technologies do not directly provide 3D structures; instead they typically yield noisy and erroneous distance information between specific entities such as atoms or residues, which have to be translated into consistent 3D models.

Here we present an approach for this translation process based on maxent-stress optimization. Our new approach extends the original graph drawing method for the new application's specifics by introducing additional constraints and confidence values as well as algorithmic components. 
Extensive experiments demonstrate that our approach infers structural models (\iec sensible 3D coordinates for the molecule's atoms) that correspond well to the distance information, can handle noisy and error-prone data, and is considerably faster than established tools. Our results promise to allow domain scientists nearly-interactive structural modeling based on distance constraints. \\[0.5ex]
\end{abstract}


\section{Introduction}
%
\textbf{Context.}
Proteins are biomolecular machines that fulfill a large variety of tasks in living systems, be it as reaction catalysts, molecular sensors, immune responses, or driving muscular activity~\cite{voet2010biochemistry}. Knowing a protein's 3D structure is a requirement for any detailed understanding of its function, and functional or structural disorder can lead to disease. Structure resolution techniques have made strong progress in recent years: biomolecules that were inaccessible a decade ago can now be structurally resolved, as exemplified by the rapid growth of structural databases~\cite{PDB2015}. The resolution techniques, however, do not directly provide structural information as 3D coordinates. Instead, \egc X-ray crystallography yields a diffraction pattern which has to be translated into a structural model. Similarly, Nuclear Magnetic Resonance (NMR) techniques measure coupled atomic spins, which can be translated into pairwise distances between specific atoms. These distances are typically incomplete, \iec not all spatially adjacent atom pairs are detected~\cite{stothers2012carbon}. Also computational tools, such as co-evolutionary analysis of multiple-sequences alignments of protein~\cite{schug2009,uguzzoni2017,ovchinnikov2017} or RNA families~\cite{leonadis2015} can provide distance constraints between residues for biomolecular structure prediction. Conceptually, one can understand the information provided by these techniques as incomplete and erroneous parts of the complete distance matrix. 

\textbf{Motivation.} Our goal is to provide an efficient and effective method to compute a full structural 3D model from incomplete and/or noisy distances. For this task, physics-based approaches are computationally prohibitive. Molecular dynamics-based approaches require weeks on supercomputers~\cite{lange2008} and stochastic global optimization techniques~\cite{schug2003} still days on medium-sized clusters. To lower computational costs, one can use more coarse-grained force-fields~\cite{rohl2004}, yet the computations still require hours to days depending on the input size. 

For interactive or nearly-interactive work, however, running times of a few seconds would be desirable. This would support a quick back-and-forth between, \egc assigning NMR chemical shifts to determine pairwise atomic distances and follow-up structural modeling\cite{stothers2012carbon}. These structural models allow then, in turn, improved NMR shift assignment and a repetition of the loop until no further improvement can be found. One forgoes describing the detailed physics and finds a near-optimal solution which respects all distance constraints. Exemplary tools which solve this distance geometry problem are \DGSOL~\cite{more1999} and \DISCO~\cite{martin2011}. Challenges for such tools are efficiency, solution quality, and support for variations of the original problem such as the ability to deal with noise, error, or distance intervals.

\textbf{Outline and Contribution.}
In this paper we transfer a maxent-stress optimization approach from 2D graph drawing~\cite{gansner2013} to computing a 3D model from (incomplete) distance information.
We exploit the resemblance of the objective functions (see Section~\ref{sec:prelim}), and extend the basic model and algorithm shown in Section~\ref{sub:maxent} by specifics of the biological application.
Our algorithmic adaptations and extensions, as well as details regarding their implementation, are presented in Section~\ref{sec:algo-impl}.

Extensive experiments (see Section~\ref{sec:experiments}) reveal that our algorithm is significantly faster than other competitive algorithms; at the same time its solution quality is very often better than the results of the best competitor. In particular in our most realistic instances, we outperform our competitors (i) in terms of quality by providing  more accurate structural models in (ii) consistently high agreement with reference models and requiring (iii) only about $5$-$10$\% of the running time. This stays true even when provided with noisy input data. 
%
We further extend our algorithm to support a weighted problem variant that allows to specify how certain a distance interval is
and obtain very promising experimental results for this setting as well. To our knowledge, our algorithm is the first to support this new variant. 

\section{Preliminaries}
\label{sec:prelim}

\subsection{Problem Definition} 
%
We model a biomolecule as a graph $G = (V, E)$, where the set $V$ of $n$ vertices models the atoms and the set $E$ of $m$ edges their relations. Distance information is given separately for all pairs $\{u,v\} \in S \subseteq V \times V$
by a distance matrix $D \in \mathbb{R}_{\geq 0}^{n \times n}$.
For this purpose $d_{vw}$ denotes the distance between vertices $v$ and $w$---or is set to \texttt{nil} if the distance is unknown (for pairs $\notin S$).
We are interested in finding an embedding of $G$ into $\mathbb{R}^3$, \iec 3D coordinates for the vertices, that respects the distances for $S$. This problem is known as (3D) \emph{distance geometry problem} (DGP).
In line with previous work, we account for inexact distance information due to measurement errors by
introducing an interval in which the actual distance is contained. This modified DGP is called \emph{interval distance geometry problem}~\cite{liberti2014survey}:
%
\begin{definition}
[Interval Distance Geometry Problem (iDGP)]
	Let a simple undirected graph $G = (V,E)$, a distance interval function $d = [l,u]$ with $d: E \rightarrow \mathbb{R}^2$, and an integer $k > 0$ be given. Then determine whether there is an embedding $x : V \rightarrow \mathbb{R}^k$ such that
	\begin{equation}
	    \label{eq:iDGP}
		\forall \{v,w\} \in E: l_{vw} \leq \lVert x_v - x_w \rVert \leq u_{vw},
	\end{equation} 
	where $l_{vw}$ and $u_{vw}$ are lower and upper bounds for the distance of the edge $\{v,w\}$. 
\end{definition}

Here and in the following, $k$ equals $3$. Then DGP is prefixed by an 'M' for 'molecular'. Note that the (M)DGP is contained in the i(M)DGP by setting the lower and upper bound of each interval to be equal to the actual distance.
%
Saxe~\cite{saxe1980} showed that deciding whether a valid embedding exists (in the DGP sense) is strongly NP-complete for $k=1$ and strongly NP-hard for $k > 1$.
Interestingly, the problem becomes solvable in polynomial time if all distances are given~\cite{blumenthal1953dg,crippen1988distance,dong2002linear}.
%
Since solving the decision problem (finding a valid embedding) is difficult and even not always possible, we continue by considering the embedding task as an optimization problem, to be solved by heuristics.
 As a measure of error, one could use the \emph{largest distance mean error} (LDME) defined as: 
 \begin{equation} \label{eq:ldme}
 \textnormal{LDME}(x) = \sqrt{\frac{1}{|E|} \sum_{ \{v,w\} \in E} \max(l_{vw} - \lVert x_v - x_w \rVert, \lVert x_v - x_w \rVert - u_{vw} , 0)^2}.
 \end{equation}
 An embedding $x$ that has an LDME value of $0$ is obviously a solution of the iDGP as each distance constraint is met. One could thus minimize the LDME of the embedding found. 

To be closer to the actual biophysical application and real-world data, we actually use the \emph{root mean square deviation} (\rmsd) to \emph{evaluate} our solutions. The \rmsd compares the embedding against a reference structure:
\begin{equation}
\label{eq:rmsd}
\textnormal{\rmsd}(x, x') =	\min \sqrt{\frac{1}{|V|} \sum_{v \in V} \lVert x_v - x'_v \rVert^2},
\end{equation}
with $x_v$ and $x_v'$ being the coordinates of the embedding and the reference structure, respectively. The minimum value is over all possible spatial translations and rotations of both superimposed structures. \rmsd values $<1.5$\AA~ approach the structure resolution limit of experimental wet-lab techniques (NMR and X-ray)~\cite{voet2010biochemistry}.
While error functions that test the capability of the algorithm to match the constraints can be useful from an optimization point of view, a good value does not necessarily mean that the embedding  reproduces the molecular structure. In particular, the algorithm must be able to handle noisy and erroneous data. In the end, the structure must also be physically meaningful. The \rmsd addresses this challenge and is a standard measure of (dis)similarity in structural molecular biology by directly assessing the usefulness in a real-life scenario\cite{voet2010biochemistry}. We will therefore rely on the \rmsd to compare algorithms.

\subsection{Related Work}
\label{sub:rel-work}
There exist many algorithms for solving the MDGP optimization problem. Yet, for most algorithms the required running time is either very high or the solution quality is in the meantime rather low. Also, some algorithms currently do not support iMDGP---which limits their use in a real-world scenario.
Due to space constraints the description of related work focuses on two tools, \DGSOL and \DISCO, which we use in our experimental comparison as they are publicly available and established in the distance geometry research community -- cf.\ a book on distance geometry edited by Mucherino \etal~\cite{mucherino2012distance} and an even more recent survey by the same authors~\cite{liberti2014survey}.
Mentioned running times (in seconds) are based on experimental data in previous works and are thus not necessarily completely comparable. For a broader overview we refer to the aforementioned book and survey.

Liberti \etal~\cite{liberti2014survey} found in their survey from 2014 that general-purpose global optimization solvers like Octave's \emph{fsolve} or spatial branch-and-bound techniques are not able to solve the problem for more than 10 vertices in a reasonable amount of time. One reason is that the objective function has a large number of local minima.
%
%

Mor\'{e} and Wu~\cite{more1997} implemented an algorithm called \DGSOL\footnote{publicly available at \url{http://www.mcs.anl.gov/~more/dgsol/} (accessed on April 4, 2017)} that transforms the objective function gradually into a smoother function that approximates the original function and has fewer local minima. The algorithm builds a hierarchy of increasingly smooth functions by iteratively applying a transformation. In a next step, \DGSOL employs Newton's method as a local optimization on the smoothest function and then traces this solution back to the original objective function, applying a local optimization on each level.  
Liberti \etal state that the algorithm ``has several advantages: it is
efficient, effective for small-to-medium-sized instances, and, more importantly, can be naturally extended to solve
iMDGP instances''~\cite[p.~23]{liberti2014survey}. On the downside, on large-scale instances the solution quality decreases (while the running time remains reasonable). 
An et al.~\cite{an2003large, an2003solving} use a different continuation approach which improves the solution quality compared to \DGSOL. In their experiments the running time of their algorithm for proteins larger than \numprint{1500} atoms lies between \numprint{500} and \numprint{1200}s.
The double variable neighborhood search with smoothing (DVS) algorithm by Liberti et al.~\cite{liberti2009double} combines the ideas of \DGSOL and variable neighborhood search into one algorithm. In their comparison to \DGSOL the quality of DVS was significantly better, but the running time two orders of magnitude slower already for small inputs.

Biswas \etal~\cite{biswas2008distributed} proposed the DAFGL algorithm that decomposes the graph into clusters by running the symmetric reverse Cuthill-McKee algorithm on the distance matrix. The subproblems are solved with a semidefinite programming (SDP) formulation. DAFGL is capable of solving the iMDGP. While its solution quality are mostly reasonable, one has to consider that as much as 70\% of distances smaller than 6~\angstrom~were provided with added noise. Also, larger instances increase the running time rapidly. The two largest molecules (PDB: 1toa and 1mqq, see Table~\ref{tab:proteins}) took already  \numprint{2654}s and \numprint{1683}s with only modest to poor solution quality (\rmsd: $3.2$\AA ~and $9.8$\AA) to solve, even though the algorithm makes use of a distributed SDP solver.
Leung and Toh~\cite{leung2009sdp} proposed the \DISCO\footnote{publicly available at \url{http://www.math.nus.edu.sg/~mattohkc/disco.html} (accessed on April 4, 2017)} algorithm that is an advancement to the DAFGL algorithm by Biswas \etal~\cite{biswas2008distributed}. 
If the problem is small enough, \DISCO solves the problem with an SDP approach and refines the obtained solution with regularized gradient descent. Otherwise, \DISCO splits the graph into two subgraphs and solves the problem recursively. \DISCO uses the symmetric reverse Cuthill-McKee algorithm to cluster the vertices initially. In a second step \DISCO tries to minimize the edge cut between different subgraphs by placing a vertex $v$ into the subgraph where most of its neighbors are placed. The algorithm puts some vertices in both subgraphs (overlapping atoms) to later stitch the two embedded subgraphs together. 
Its authors tested \DISCO also in the iMDGP setting (\iec \DISCO supports inexact distances). The results indicate that \DISCO is able to compute the structure of proteins with very sparse distance data and high noise in \numprint{412}s for a protein having \numprint{3672} atoms. 
Fang and Toh~\cite{fang2013using} presented some enhancements to \DISCO for the iMDGP setting by incorporating knowledge about molecule conformations to improve the robustness of \DISCO. Their experiments show that their changes indeed lead to better solutions (about 50--70\%) with the cost of increased running times (also about 50--70\%).
%

\subsection{MaxEnt-Stress Optimization}
\label{sub:maxent}
We aim at developing an algorithm for iMDGP (and its extension wiMDGP, see Section~\ref{sub:algo-wiMDGP}) with a significantly lower running time than previous algorithms and with solutions of good quality.
%
Our main idea is to use an objective function proposed by Gansner \etal~\cite{gansner2013} for planar graph drawing, called maxent-stress (short for \textit{maximal entropy stress}). As the name suggests, it is composed of two parts, a stress part that penalizes deviations from the prescribed distances (with a quadratic penalty, possibly weighted) and an entropy part that penalizes vertices for getting too close to each other (atoms cannot overlap):

\begin{equation} \label{eq:maxent-stress}
	\min_x \sum_{vw \in S} \omega_{vw} (\lVert x_v - x_w \rVert - d_{vw})^2 \; - \alpha H(x),
\end{equation} 
where $H(x) = -\operatorname{sgn}(q) \sum_{vw \notin S} \lVert x_v - x_w \rVert ^{-q}, q > -2$, is the entropy term, 
$\omega_{vw}$ a weighting factor for edge $\{v,w\}$, $\alpha \geq 0$ a user-defined control parameter, and $\operatorname{sgn}(q)$ the signum function with the special case that $sgn(0) = 1$. 

To minimize function~(\ref{eq:maxent-stress}), Gansner \etal~\cite{gansner2013} derive a solution from 
successively solving Laplacian linear systems of the form $Lx=b$ for $x$. A noteworthy feature of this successive iteration towards a local minimum is that the solution of the current iteration depends on the solution of the previous iteration, since the current right-hand side is computed from the solution in the previous iteration.
Note that the computation of distances between vertex pairs not in $S$ is not required for function~(\ref{eq:maxent-stress}). Instead, vertex pairs not in $S$ are related to each other via the entropy term, which enters the right-hand side, too. If the parameter $q$ is set to be smaller than zero, the entropy term of vertex $u$ acts as a sum of attractive forces on $u$. Conversely, the term acts as a sum of repulsive forces if $q$ is larger than zero.

The Gansner \etal algorithm typically needs several iterations to converge. In this process the entropy weighting factor $\alpha$ has a strong influence in the maxent-stress model: a high value will cause the vertices to expand into space indefinitely while a low value will cause no entropy influence. The maxent-stress algorithm therefore starts with $\alpha = 1$ and gradually reduces this value to $\alpha = 0.008$ with a rate of $0.3$. For each value of $\alpha$, a maximum of 50 linear system solves are performed and Gansner \etal set $q$ to $0$ except when the graph has more than 30\% degree-1 vertices (then $q \gets 0.8$). 
Note that in this entropy context they assume $\Vert x \Vert^0 = \ln \Vert x \Vert$.
If the relative difference $\lVert x' - x \rVert / \lVert x \rVert$ between two successive solutions $x$ and $x'$ is below $0.001$, the algorithm is terminated.  

More implementation details (\egc the approximation of the entropy term in case of $|S| \in \bigO{n}$) can be found in Section~\ref{sec:impl}.

%


\section{New Algorithm and its Implementation}
\label{sec:algo-impl}
Now we describe the adaptations made to the generic maxent-stress algorithm in order to deal with iMDGP and an extension called wiMDGP. For more technical details we refer the interested reader to our code\footnote{\url{https://algohub.iti.kit.edu/parco/NetworKit/NetworKit-MaxentStress}, main source folder \texttt{networkit/cpp/viz}. An updated
    standalone version is planned to be published in the future.}.

\subsection{Adapting the Maxent-stress Algorithm for iMDGP}

Recall that for iMDGP we are given a graph $G = (V, E)$ and the distance intervals $d = [l,u] : E \rightarrow \mathbb{R}_{\geq 0}^2$. We then want to find an embedding $x: V \rightarrow \mathbb{R}^3$ that respects the intervals as well as possible. Note that, in line with previous work, we assume the set $S$ of known distances to be equal to $E$ here. We also assume the edges in $E$ to be unweighted.

As Gansner \etal's maxent-stress algorithm~\cite{gansner2013} cannot cope with intervals, we solve the iMDGP by first running our implementation of the maxent-stress algorithm with an adapted distance $d' : E \rightarrow \mathbb{R}$ that is defined by $d'_{vw} \coloneqq (l_{vw} + u_{vw}) / 2$. One might expect that, in the resulting embedding, the distances should be roughly in the middle of their interval. This would already be a valid solution to the iMDGP. However, our preliminary experiments show that the output of the maxent-stress algorithm still violates distance constraints even on smaller graphs if called this way. We therefore apply local optimizations to the layout computed by the maxent-stress algorithm. One optimization is based on simulated annealing, while the other is a simple local optimization algorithm.

\paragraph*{Optimization of the Embedding with Simulated Annealing.}
Simulated annealing~(SA) is a well-known metaheuristic that can escape local optima by probabilistically (based on a temperature parameter) accepting neighbor solutions that are worse than the current one, see \eg Talbi~\cite{Talbi:2009:MDI:1718024}.

Our SA algorithm is sketched as Algorithm~\ref{alg:sa_optimizer} in Appendix~\ref{sub:sa-alg}. 
The SA metaheuristic is often especially powerful if the initial solution is randomly chosen and can then jump between local minima. In our setting we receive an embedding from the maxent-stress algorithm as input; its global structure should be already quite good and only some of the given distances are not in their desired intervals. Therefore, our SA algorithm is only used to overcome rather narrow local minima instead of jumping to a completely different solution. 

The constants in Algorithm~\ref{alg:sa_optimizer} have been manually chosen in informal experiments.
The outermost loop breaks after a certain number of unsuccessful improvement attempts or if the temperature is really low. The second loop iterates until an equilibrium \wrt the current temperature is reached---here controlled by the number of iterations and modifications.
As we do not want to get completely different solutions for reasons mentioned above, we choose a low start temperature and decrease it rather quickly.

Within the innermost loop a new neighbor solution is computed. In fact, we use parallelism here to reduce the running time of the algorithm. While this can lead to some data races if the position of a vertex is altered by more than one thread, we did not experience any significant decrease in quality. The number of total iterations, steps with no improvement and the number of modifications are all chosen rather small to keep the running time low.

The main ingredients of the innermost loop are (i) the local error criterion, (ii) the local optimizer that computes a neighbor solution, and (iii) the acceptance function.

We define $\text{error}_{vw}(x)$ as  
$\text{error}_{vw}(x) \coloneqq \max \{ l_{vw} - \lVert x_v - x_w \rVert, \lVert x_v - x_w \rVert - u_{vw} , 0\}^2.
$
and the local error of an edge $\{v,w\}$ as:
\[
\text{localError}(\{v,w\}, x) \coloneqq \text{error}_{vw}(x) + \sum_{\mathclap{u \in N(v) \setminus \{w\}}} \text{error}_{vw}(x) + \sum_{\mathclap{u \in N(w) \setminus \{v\}}} \text{error}_{uw}(x),  
\]
where $N(v)$ denotes the neighborhood (\ie the set of incident vertices) of $v$.

In each iteration a new neighbor solution is computed for an edge $\{v,w\}$.
To this end, we apply a force-based approach that takes the edge lengths to their neighbors and the edge length of $\{v,w\}$ into account. 
The idea is to model the local system similarly to the spring embedder model~\cite{eades1984} and the force-directed algorithm by Fruchterman and Reingold~\cite{fruchterman1991}. The difference to those algorithms is that we have to deal with an interval for the length of an edge. In our local force optimization step, we only change the positions of $v$ and $w$ while keeping adjacent vertices fixed. 

We say an edge $\{v,w\}$ is \emph{violating} its distance constraint if the error$_{vw}(x)$ is larger than $10^{-9}$ (and not exactly 0 due to numerical reasons). In our spring model only the violating edges account for a repulsive or attractive force, while the other edges do not take part in the force calculation. In our model each spring has its equilibrium state in an interval that corresponds to the interval of the respective edge it models.

For an optimization on edge $\{v,w\}$, the forces acting on $v$ and $w$ are a combination of attractive and repulsive forces: $f(v) \coloneqq f_{rep}(v) + f_{attr}(v)$ and $f(w) \coloneqq f_{rep}(w) + f_{attr}(w)$. The repulsive and attractive forces for a vertex $v$ are defined as 
$f_{rep}(v) \coloneqq \sum_{w \in N_{rep}(v)} (x_v - x_w) \cdot \frac{l_{wv}^2}{\lVert x_w - x_v \rVert^2}$
and
$f_{attr}(v) \coloneqq \sum_{w \in N_{attr}(v)} (x_w - x_v) \cdot \frac{u_{wv}^2}{\lVert x_w - x_v \rVert^2},$
where $N_{rep}(v) \subseteq N(v)$ is the set of neighbors of $v$ that are too close to $v$ (\iec the edge is shorter than its lower bound) and $N_{attr}(v) \subseteq N(v)$ is the set of neighbors of $v$ that are too far away (\iec the edge is longer than its upper bound). 

Finally, the acceptance function always accepts improving changes. Error-increasing changes are probabilistically accepted according to the Boltzmann distribution based on the local error (as in many cases~\cite{Talbi:2009:MDI:1718024}).

\paragraph*{A Simple Local Optimization Algorithm.}

In addition to our SA optimization algorithm, we propose another simple local optimization algorithm. During one iteration the algorithm sorts the edges by their error (\iec the deviation from the edge's given distance interval) in descending order. For an edge $\{v,w\}$ having a length that is not in the given distance interval, the algorithm either prolongates or shortens the edge length such that it is exactly as long as the upper or lower bound given by the distance interval, respectively. We only accept the change if we reduce the local edge error.  If we change the length of an edge $\{v,w\}$, we lock the other incident edges of $v$ and $w$ for the remainder of the current iteration to prevent an oscillating effect. We perform a maximum of \numprint{50} iterations or less if there is no improvement between two successive iterations. Pseudocode of the method is shown as Algorithm~\ref{alg:simple-local} in Appendix~\ref{sub:simple-local}.

\subsection{Intervals with Confidence Values: wiMDGP}
\label{sub:algo-wiMDGP}
Some distances can be measured more accurately than others in common biomolecular experimental methods.
To account for this, we add a confidence to each interval. Such a confidence states how certain it is that the actual distance is contained in this interval,
leading us to the following problem definition:
\begin{definition}
[Weighted Interval Distance Geometry (Optimization) Problem (wiDGP)]
Let a simple undirected graph $G = (V,E)$, a distance interval function $d = [l,u]$, a confidence function $p: E \rightarrow \mathbb{R}$, and an integer $k > 0$ be given. Then minimize the following function:
	\begin{equation}
    \label{eq:icDGOP}
		\sum_{\{v,w\} \in E} \omega_{vw} \cdot \text{error}_{\{v, w\}}(x),
	\end{equation} 
	where the weight $\omega_{vw}$ depends on the edge's confidence value $c_{vw}$.
\end{definition}

In order to support wiMDGP, we adapt the maxent-stress algorithm as well as the other two optimization algorithms. 
For wiMDGP we can use the weights $\omega_{vw}$ from Eq.~(\ref{eq:maxent-stress}), the maxent-stress optimization problem, as a penalty that increases the error of an edge if the confidence that the distance lies in the interval is high. After some manual parameter tuning, we have chosen the following function to define the weighting factors:
$\omega_{vw} \coloneqq 1 + 5\exp^{-5 (1 - c_{vw})},$
%
where $c_{vw}$ denotes the confidence for edge $\{v,w\}$. This way, the weight of an edge is roughly in the interval $[1, 6]$ and increases rapidly between \mbox{$0.7$ and $1$.} A confidence between $0$ and $0.6$ only tells us that we cannot be very certain about the distance and thus, the error term should not vary too much. For higher confidence values, we can be quite certain that the distance interval is correct, so we need to penalize the errors of such edges significantly higher. Choosing a larger interval for the weights turned out not to be beneficial in terms of solution quality in preliminary experiments.


Both our SA and simple local optimization algorithm use an adapted error function for an edge that includes the weighting function above. In our simple local optimizer, we additionally change the sorting of the edges such that edges with higher confidence are considered first. Confidence ties are broken by choosing the edge with larger error first.

\subsection{Implementation Details}
\label{sec:impl}
%
\subparagraph*{Initial layout.}
For computing the initial layout, we implemented three algorithms. In addition to PivotMDS~\cite{brandes}, which was used by Gansner \etal~\cite{gansner2013}, these are two random vertex placement algorithms.
The first one is a very simple method and places the coordinates randomly in a $k$-dimensional hypercube with predefined side length.

The second one does include some of the distance information provided by the input. 
Given an edge $\{v,w\}$ and the coordinates of vertex $v$, we place $w$ at the boundary of a $k$-dimensional hypersphere with radius $d_{vw}$ and centered at $v$. The algorithm needs a start vertex with given coordinates, as neighbors will be placed around it. We choose the vertex with maximum degree in the graph; this way, the highest number of neighbors is placed in random directions around the start vertex, which should lead to better spatial spread.

Finally, PivotMDS is an approximation algorithm for multidimensional scaling that is based on sampling; for a detailed description the reader is referred to Brandes and Pich~\cite{brandes}.

Preliminary experiments indicated that PivotMDS and the random hypersphere approach fare similarly well. Since PivotMDS turned out to be more robust in terms of solution quality when applied to protein instances, we use it in all the following experiments. Its slower speed is more than outweighed by the more expensive maxent-stress algorithm.

\subparagraph*{Approximating the entropy term.}
The entropy term in Eq.~(\ref{eq:maxent-stress}) iterates over all elements not in $S$. As the set $S$ is usually sparse, this computation would thus require quadratic running time. Thus, it is important to approximate the distances required for the entropy calculations. We implemented both the well-known Barnes-Hut approximation (also used by Gansner \etal) as well as well-separated pair decomposition (WSPD)~\cite{callahan1995}. 

Additionally, we evaluate the entropy lazily:
instead of computing the entropy term in each iteration, we recompute it only when the function $\lfloor 5 \log i\rfloor$ changes, where $i$ is the iteration number. 
We expect the entropy term to significantly change more frequently at the beginning of a new iteration; thus, we use a function that causes the algorithm to recompute the entropy more often at lower iteration numbers. 
Lipp \etal~\cite{lipp2015} use the same function for reducing the running time of their WSPD algorithm.
This ``lazy'' computation of the entropy term significantly reduces the running time while the quality does not deteriorate much.

\subparagraph*{Solving the Laplacian linear systems.}
Recall that Gansner \etal derived an iteration of subsequent Laplacian linear systems for optimizing maxent-stress. They use the conjugate gradient method (CG) as a Laplacian solver in their implementation.
The conjugate gradient method has superlinear time complexity. That is why we use lean algebraic multigrid (LAMG) instead, a fast solver proposed by Livne and Brandt~\cite{livne2012} with linear empirical running time.
We use our C++ implementation of LAMG; it is available in NetworKit and has been used for other Laplacian graph problems before~\cite{DBLP:conf/siamcsc/BergaminiWLM16}. An alternative multilevel approach for solving the linear systems exists~\cite{meyerhenke2015}, but is harder to adapt to the present scenario.

\subparagraph*{Optimization Workflow.}
We combine our two optimization algorithms into one workflow: the solution found by the SA algorithm can often be further improved by a subsequent run of our simple local optimization algorithm. Sometimes it happens that the SA algorithm only finds a slightly worse solution compared to the maxent-stress algorithm. In this case we ignore the SA solution and only run the simple optimization algorithm.

\section{Experiments}
\label{sec:experiments}
In this section we present a representative subset of our experiments and their results. To this end, we implemented our algorithm in C++ based on NetworKit~\cite{staudtSM14}, an open-source toolkit for graph algorithms and in particular interactive large-scale network analysis. We call our algorithm \NWK (for \textbf{M}axent-stress \textbf{O}ptimization of \textbf{Bi}omolecular models) and compare it with \DGSOL~\cite{more1997} (C/Fortran code) and \DISCO~\cite{leung2009sdp} (compiled Matlab code), two of the very few publicly available established tools that can handle inexact inputs with intervals.

All experiments are done on a machine that has 256 GB of main memory and 2 sockets, each equipped with an 8-core Intel\textregistered~Xeon\textregistered~CPU E5-2680 clocked at 2.7 GHz (hyperthreading enabled). We compile our code with g++ version 4.8.1 and OpenMP 3.1. The MATLAB version installed on the system is MATLAB\textregistered~R2015a (8.5.0.197613) 64-bit.

\subsection{Instances}
\label{sub:exp-settings}
To test the accuracy and efficiency of our approach in a real-world setting, we work on 10 proteins of different sizes~\cite{berman2000} taken from the protein data bank (PDB)~\cite{PDB2015}, see Table~\ref{tab:proteins}. These proteins range from small globular proteins with 50 amino acids to large proteins with about 700 amino acids. We only consider \emph{ATOM} entries in the file, which provide atomic coordinates. Also, we only work on the first chain in the case of multiple protein chains. Based on the coordinates of a protein, we can construct an instance for the various distance geometry problems.
For each experiment we actually create three instances per protein (\iec different contact distance information) to eliminate effects of particularly good/bad sets of input data. Also, each instance (\iec same contact distance information) is re-run three times to eliminate stochasticity effects. The displayed data per protein are averaged over these three instances and three respective runs.

\begin{table}[tbh]
	\caption[Proteins for distance geometry benchmarks]{Proteins for distance geometry benchmarks and their basic properties~\cite{berman2000}. Listed are the protein data bank (PDB) code~\cite{PDB2015},  and the size in atoms (vertices), amino acid residues, the number of edges equivalent to covalent bonds (= bonds edges) and the number of edges with atoms closer to each other than 5\AA~without being a covalent bond~(= contact edges). 
    }
	\label{tab:proteins}
	\centering
	\nprounddigits{2}
	\begin{small}
   \begin{tabular}{r r r r r}
        \hline
            Protein & \# atom/ vertices & \#  residues & bond edges & contact edges\\
        \hline
            1ptq &  402 &  50 &  412 &  3987\\
            1lfb &  641 &  78 &  654 &  6320\\
            1gpv &  735 &  87 &  696 &  7208\\
            1f39 &  767 & 101 &  788 &  7621\\
            1ax8 & 1003 & 130 & 1016 & 10527\\
            1rgs & 2015 & 264 & 2053 & 20731\\
            1toa & 2138 & 277 & 2181 & 23168\\
            1kdh & 2846 & 356 & 2904 & 30655\\
            1bpm & 3671 & 481 & 3744 & 41027\\
            1mqq & 5510 & 675 & 5665 & 62396\\
        \hline
    \end{tabular}
	\end{small}
\end{table}

We use a percentage $p$ of all atom distances $\{v,w\} \in \binom{V}{2}$ for which $d_{vw} = \lVert x_v - x_w \rVert$ is below a cut-off distance of 5 \AA, where $x$ denotes the coordinates from the protein file. We use 5~\AA~because this approximates the distance that can be determined by NMR experiments~\cite{wuthrich1990,stothers2012carbon} and since it is a typical cutoff in determining so-called contact maps (\ie binary matrices that store only adjacencies whose length is below the cutoff)~\cite{schug2010,noel2013}. To construct an instance for iMDGP, we introduce the interval $[d_{vw} - \underline{\epsilon}, d_{vw} + \overline{\epsilon}]$, where $\underline{\epsilon}, \overline{\epsilon} = d_{vw} \cdot \mathcal{N}(0, \sigma^2)$ and $\sigma$ is the standard deviation. We denote instances created this way as \mbox{\emph{normal}-iMDGP} instances.

In contrast, the \mbox{\emph{bonds}-iMDGP} more closely reflects standard NMR experiments by taking protein biochemistry into account. As all covalent bonds of a protein are known, instances can be assumed to have full knowledge of exact distances for these bonds: Chemically, the length of covalent bonds fluctuates very little, hence the interval for these bonds edges $\{v,w\}$ consists of a single distance $[d_{vw}, d_{vw}]$. In addition to the distance information of the bonds, we add more distances the same way as for constructing a \mbox{\emph{normal}-iMDGP} instance.

\subsection{Results \& Discussion}
\label{sub:exp-results}
To quantify the structure determination quality of the different approaches, we compare them by  \rmsd. For the \mbox{\emph{normal}-iDGP} test instances, we observe low \rmsd for \NWK and \DISCO, while \DGSOL performs significantly worse (cf.\ Tables~\ref{tab:t50_0.1} and~\ref{tab:t30_0.1} as well as Table~\ref{tab:t70_0.1} in Appendix~\ref{sub:add-exp}). As expected, the \rmsd gets higher if less distance information is provided (the instances in Tables~\ref{tab:t50_0.1} and~\ref{tab:t30_0.1} provide only 50\% and 30\% of the contact edges, respectively, while the instances in Table~\ref{tab:t70_0.1} provide 70\%). The \rmsd values do not increase, as one might expect \textit{a priori}, necessarily with instance size (number of atoms/ vertices). Instead, these instances have more edges, which might make the embedding computationally more demanding but also of good quality. Indeed, \NWK and \DISCO yield good embeddings regardless of system size. \DGSOL performs worse for larger systems. Also, \NWK is more consistent than \DISCO and performs best in nearly all instances, in particular for those with less information (cf.\ instances 1gpv and 1rgs in Table~\ref{tab:t30_0.1}). Similarly, the running times of \NWK are about an order of magnitude faster than \DGSOL and \DISCO, with \DGSOL being slightly faster than \DISCO. There is an overall  trend of increased running times with system size, but some instances seem particular hard to compute (\eg instances 1gpv, 1rgs, and 1toa in Table~\ref{tab:t30_0.1}). 
Gaussian noise on the intervals does not significantly alter the results: given a relatively high  amount of distance information, \NWK and \DISCO produce embeddings of very high quality with \rmsd $< 1$\AA\ (see Tables~\ref{tab:t70_0.1}, \ref{tab:t70_0.01} and~\ref{tab:t70_0.001} in Appendix~\ref{sub:add-exp}). It should be noted, though, that in Table~\ref{tab:t70_0.01} \DISCO yields the majority of best results in terms of solution quality---but \NWK is usually not far behind.

\begin{figure}[tbh]
\centering
 \begin{minipage}{0.36 \textwidth}
       \includegraphics[width=\textwidth]{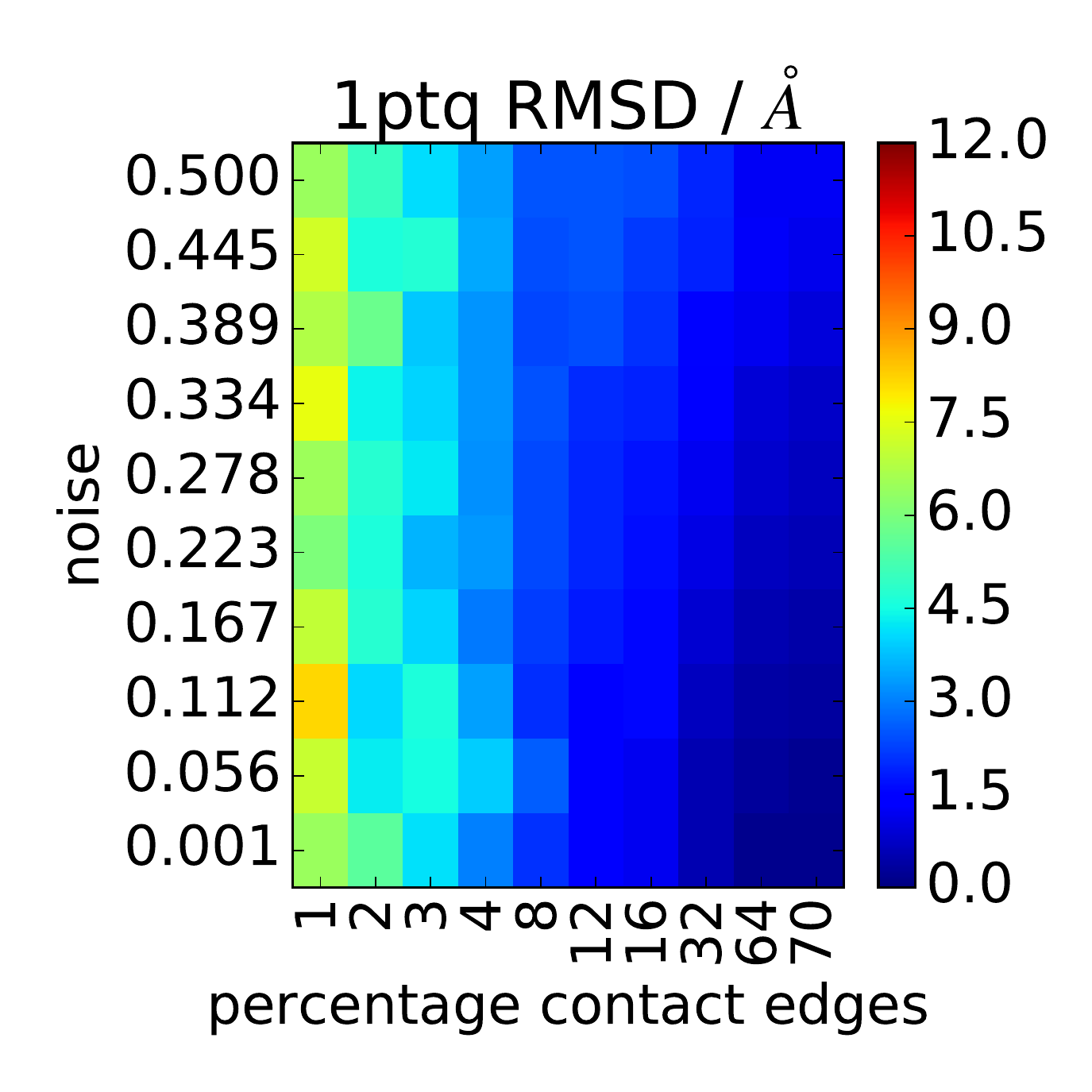}
   \end{minipage}
     \begin{minipage}{0.36 \textwidth}
       \includegraphics[width=\textwidth]{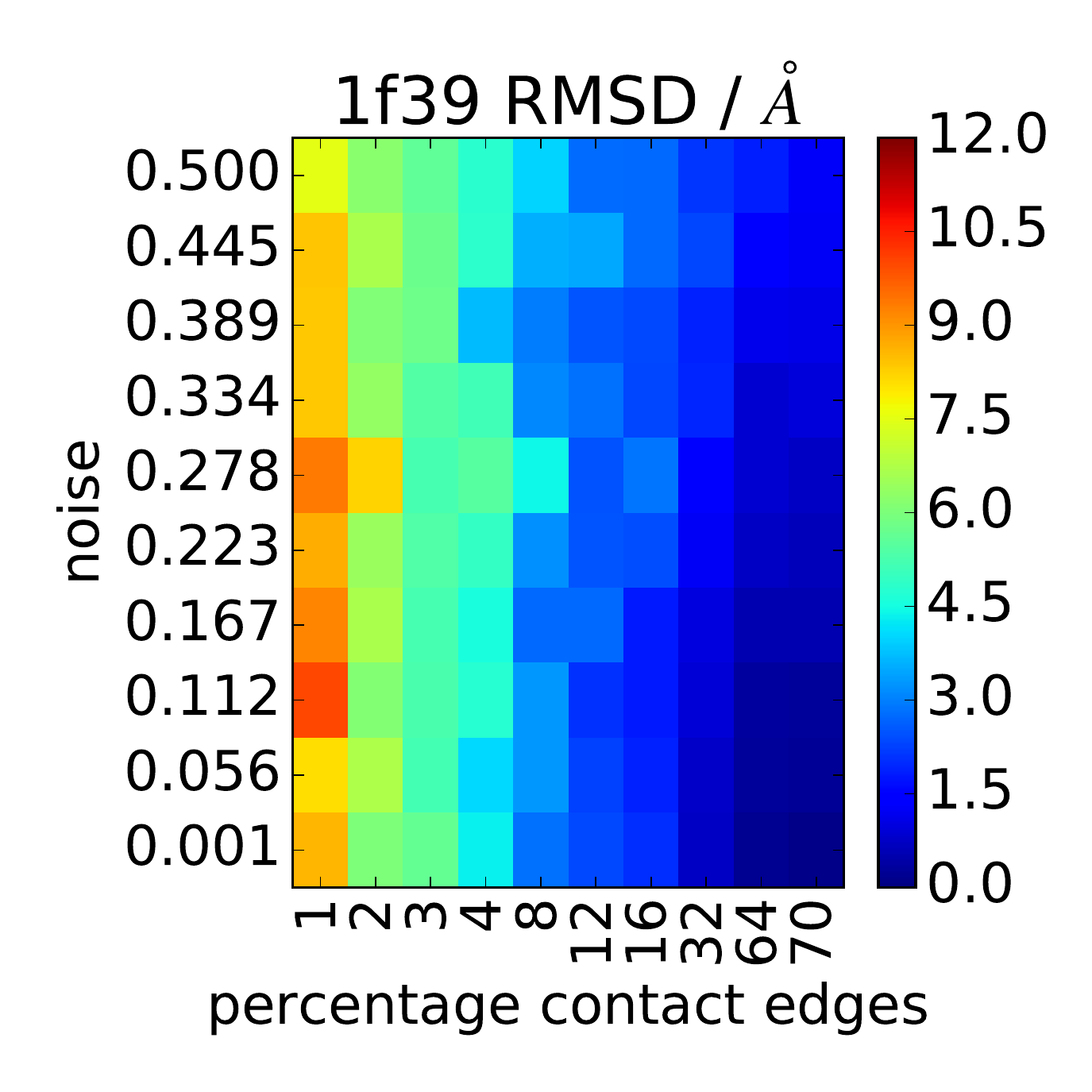} 
      \end{minipage}
      
     \begin{minipage}{0.36 \textwidth}
       \includegraphics[width=\textwidth]{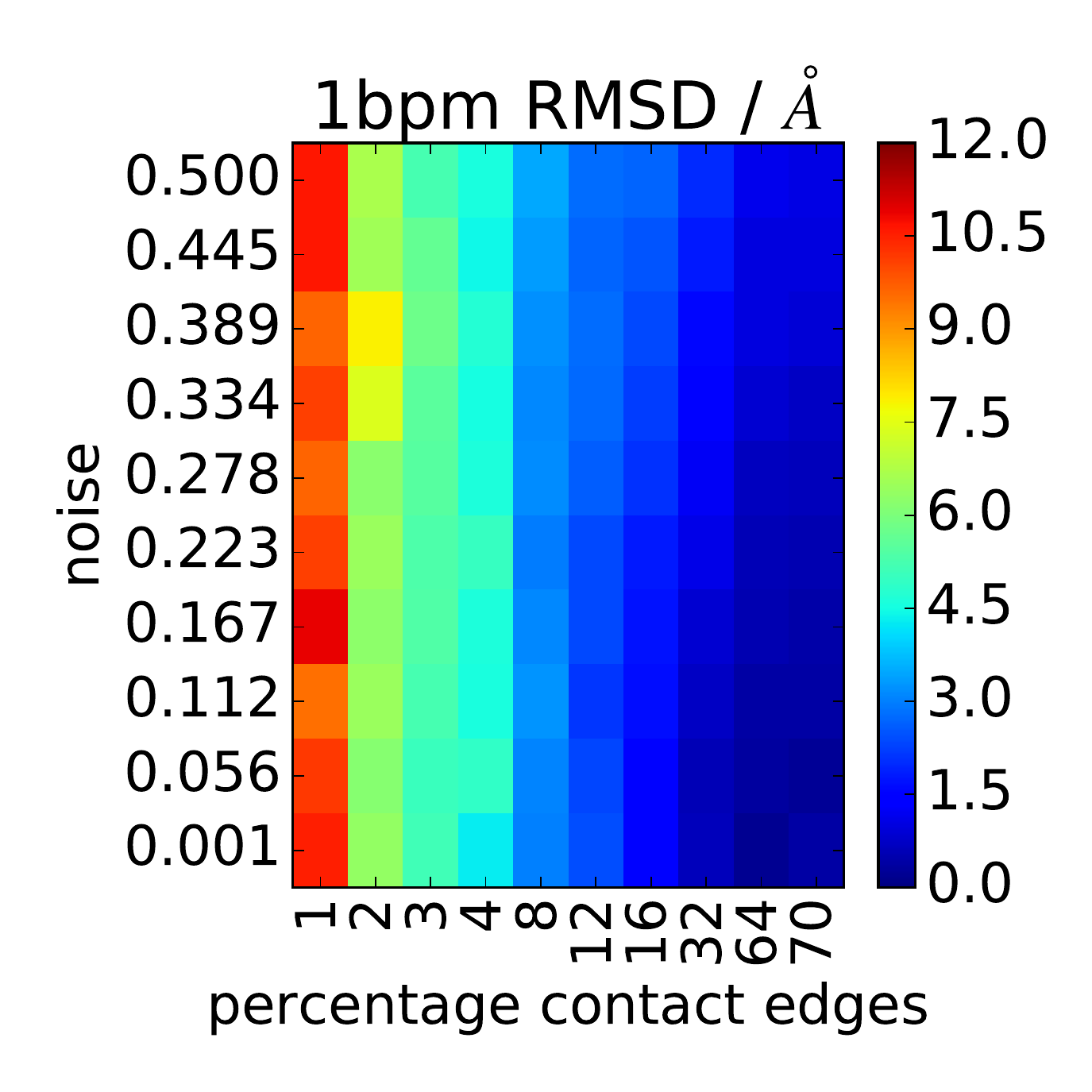} 
      \end{minipage}
     \begin{minipage}{0.36 \textwidth}
       \includegraphics[width=\textwidth]{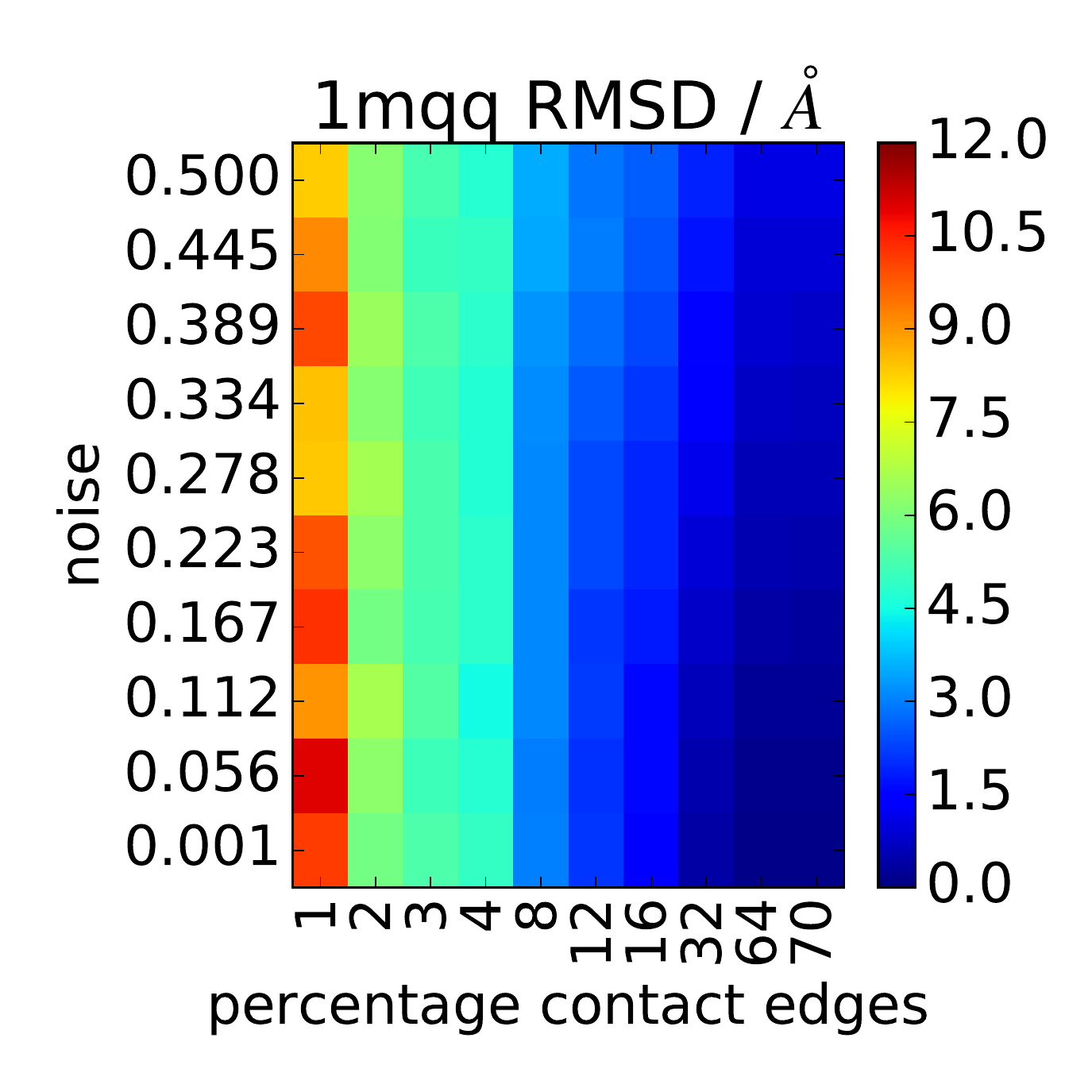} 
   \end{minipage}
   \caption{Quality results for bonds-iDGP instances. The horizontal axis shows the amount of contact edges in addition to bond edges provided, the vertical axis varies the $\sigma$ of Gaussian noise.}
   \label{heatmap_bonds}
   \end{figure}

For the \mbox{\emph{bonds}-iMDGP} instances, we display the results only for \NWK (\DISCO and \DGSOL show the same respective trends as above) as a heatmap in Figure~\ref{heatmap_bonds}. For very low amounts of additional distance information (1-2\%) in addition to the bonds, \NWK is unable to provide high-quality embeddings as displayed by \rmsd values $> 5$\AA. When provided as little as 8-12\% distance information in addition to the bonds, the structure determination quality becomes below $3$\AA\ \rmsd, \iec it approaches the wet-lab resolution. Interestingly, the amount of Gaussian noise does not strongly influence the embedding quality.  Exemplary structure embeddings are displayed in Figures~\ref{1mqq} and~\ref{1ptq} in Appendix~\ref{sub:add-exp}. While the overall quality appears good here as well, one can see that most structural errors are found at the surface, where fewer edges are available.

The experimental results for \mbox{wiMDGP} are shown in Table~\ref{tab:wiMDGP-t50} in Appendix~\ref{sec:add-exp-wiMDGP}. As these instances cannot be directly compared against other cases, we merely report that \NWK produces high quality solutions, typically with \rmsd $<1.0$\AA\  and  comparable running times to the other instances with \NWK. 
To truly assess the use of this \mbox{wiMDGP} implementation on real-world data, one would have to work on curated wet-lab experimental data~\cite{wuthrich1990,voet2010biochemistry} or co-evolutionary signals~\cite{schug2009}. This is clearly outside the scope of this paper.

Overall, in particular for limited and noisy distance information, \NWK provides consistently embeddings with higher quality and does so at significantly lower running times than both \DISCO and \DGSOL. On average (geometric means over quotients for each of the Tables~\ref{tab:t50_0.1} to~\ref{tab:t70_0.001}), \NWK is between \numprint{13}x and \numprint{20}x faster than \DISCO. At the same time its RMSD values are on average \numprint{17}\% to \numprint{41}\% better---except for Table~\ref{tab:t70_0.01}, where \DISCO is \numprint{12}\% better.
\DGSOL is not competitive in terms of solution quality and also \numprint{6}x to \numprint{13}x slower.

For \mbox{\emph{normal}-iDGP} instances and very high amounts of distance information, the \NWK embeddings provide \rmsd $<1$\AA, which is below the typical resolution of NMR or X-ray~\cite{voet2010biochemistry}; even providing only $p=30\%$  leads to high quality embeddings.  Both \NWK and \DISCO perform considerably better than older algorithms such as\cite{biswas2008distributed}, where some \rmsd $>5$\AA\ were reported. In the more realistic scenario of \mbox{\emph{bonds}-iMDGP} instances with all bond edges provided, only few contact edges (8-12\%)  can already lead to high quality embeddings with \NWK. Thus, we are confident that our algorithm will lead to improved interpretation of wet-lab experiments in particular in cases with sparse data, such as sparse NMR experiments.  

%
%
%

\begin{table}[t]
	\caption[Performance results on 50\%-0.1 \mbox{\emph{normal}-iDGP}]{Performance results on $50\%$ $\sigma=0.1$ \mbox{\emph{normal}-iDGP} instances. 
    Best results in bold font.}
        \label{tab:t50_0.1}
	\centering
	\nprounddigits{2}
    \begin{small}
	\begin{tabular}{l | r r r | r r r}
		\hline
	 	 & \multicolumn{3}{c}{\textbf{\rmsd \ / \AA}} & \multicolumn{3}{c}{\textbf{time \ / s}} \\
	 	Protein & \NWK & \DGSOL & \DISCO & \NWK & \DGSOL & \DISCO \\
		\hline
		1ptq & \numprint{0.460490707} & \numprint{8.052750333} & \textbf{\numprint{0.453339215}} &\textbf{ \numprint{1.561111111}} & \numprint{9.85111111} & \numprint{13.70333333} \\
		1lfb & \textbf{\numprint{0.868549066}} & \numprint{10.50872991} & \numprint{0.997395596} &\textbf{ \numprint{2.725555556} }& \numprint{22.99333333} & \numprint{25.28777778} \\
        1gpv & \textbf{\numprint{0.677997608}} & \numprint{14.35347047} & \numprint{0.905033569} & \textbf{\numprint{7.50888889}} & \numprint{120.3522222} & \numprint{128.5444444} \\
        1f39 & \textbf{\numprint{0.526879951} }& \numprint{16.45388764} & \numprint{0.687227121} & \textbf{\numprint{5.552222222} }& \numprint{70.21555556} & \numprint{70.80444444} \\
        1ax8 &\textbf{ \numprint{0.506903837} }& \numprint{12.22901154} & \numprint{0.552167271} &\textbf{ \numprint{3.913333334}} & \numprint{39.41666667} & \numprint{48.48777778} \\
         1rgs & \textbf{\numprint{0.600526734}} & \numprint{17.55806664} & \numprint{1.0504721} & \textbf{\numprint{7.102222222} }& \numprint{112.4911111} & \numprint{155.520} \\
		1kdh &\textbf{ \numprint{0.876699213}} & \numprint{19.40933222} & \numprint{1.059098055} & \textbf{\numprint{8.636666667}} & \numprint{173.6988889} & \numprint{186.0044444} \\
		1toa & \textbf{\numprint{0.482768238}} & \numprint{23.71987128} & \numprint{0.857911622} &\textbf{ \numprint{14.13555556} }& \numprint{181.960} & \numprint{311.923333} \\
		1bpm &\textbf{ \numprint{0.484492792}} & \numprint{21.72628603} & \numprint{0.496694857} &\textbf{ \numprint{12.93666667} }& \numprint{221.180} & \numprint{264.020} \\
		1mqq & \textbf{\numprint{0.336321992} }& \numprint{23.55773643} & \numprint{0.404776578} & \textbf{\numprint{18.21666667} }& \numprint{381.6544444} & \numprint{519.5811111} \\
		\hline
	\end{tabular}
    \end{small}
\end{table}

\begin{table}[t]
	\caption[Performance results on 30\%-0.1 \mbox{\emph{normal}-iDGP}]{Performance results on $30\%$ $\sigma=0.1$  \mbox{\emph{normal}-iDGP} instances. 
        Best results in bold font.}
        \label{tab:t30_0.1}
	\centering
	\nprounddigits{2}
    \begin{small}
	\begin{tabular}{l | r r r | r r r}
		\hline
	 	 & \multicolumn{3}{c}{\textbf{\rmsd \ / \AA}} & \multicolumn{3}{c}{\textbf{time \ / s}} \\
	 	Protein & \NWK & \DGSOL & \DISCO & \NWK & \DGSOL & \DISCO \\
		\hline
		1ptq & \textbf{\numprint{1.0495372}} & \numprint{9.28639785} & \numprint{1.088837409} & \textbf{\numprint{1.375555556}} & \numprint{12.38555556} & \numprint{13.29888889} \\
		1lfb &\textbf{ \numprint{1.497364068}} & \numprint{11.67719797} & \numprint{1.532461703} & \textbf{\numprint{2.794444445}} & \numprint{26.75666667} & \numprint{22.86333333} \\
		1gpv & \textbf{\numprint{1.293614779}} & \numprint{16.2924845} & \numprint{3.478011883} & \textbf{\numprint{6.641111111}} & \numprint{112.0777778} & \numprint{114.6344444} \\
		1f39 &\textbf{ \numprint{1.064044002}} & \numprint{17.274883} & \numprint{1.74229217} &\textbf{ \numprint{5.177777778}} & \numprint{82.75222222} & \numprint{78.07222222} \\
        1ax8 & \textbf{\numprint{1.074819516}} & \numprint{12.95360221} & \numprint{1.335558966} & \textbf{\numprint{3.847777778}} & \numprint{47.510} & \numprint{44.43111111} \\
		1rgs &\textbf{ \numprint{1.366124967}} & \numprint{18.02871975} & \numprint{5.53842692} &\textbf{ \numprint{40.330}} & \numprint{120.8866667} & \numprint{120.5933333} \\
        1toa &\textbf{ \numprint{0.998294844}} & \numprint{23.4717915} & \numprint{1.27695637} & \textbf{\numprint{13.10111111}} & \numprint{204.9444444} & \numprint{325.6444444} \\
		1kdh & \textbf{\numprint{1.461608359}} & \numprint{20.32886951} & \numprint{1.508157275} &\textbf{ \numprint{8.282222222}} & \numprint{174.6733333} & \numprint{169.410} \\
        1bpm & \textbf{\numprint{0.932015435}} & \numprint{22.08108639} & \numprint{1.368421798} & \textbf{\numprint{11.49111111}} & \numprint{237.230} & \numprint{255.8688889} \\
		1mqq & \textbf{\numprint{0.820865379}} & \numprint{23.44579063} & \numprint{0.961508028} & \textbf{\numprint{16.36111111}} & \numprint{216.3888889} & \numprint{529.0855556} \\
		\hline
	\end{tabular}
    \end{small}
\end{table}

\section{Conclusions}
\label{sec:concl}
This paper provides a significant step towards nearly-interactive protein structure determination. 
To this end, we implemented the maxent-stress algorithm~\cite{gansner2013} and incorporated first of all a faster Laplacian solver. Based on this implementation we extended the graph drawing algorithm to handle distance geometry problems where the distances are either exact or come in the form of intervals, optionally with some confidence.
Comparing our algorithm with two publicly available competitors shows that we are able to significantly outperform both of them in terms of running time, while usually providing embeddings with higher quality. For the more realistic bonds-iDGP instances, our algorithm is able to compute high quality protein structures with limited and noisy information. Most errors can be found at the surface of the proteins, where only few edges can guide the optimization process.

While some related work can provide even higher solution quality (\eg~\cite{fang2013using}), it can only do so at the expense of an enormous increase in running time. The strength of our work is the combination of low running time, good and consistent solution quality, and genericity.

The evaluation of our algorithm on real-world instances whose distance matrices are derived from chemical bonds and real NMR experiments is ongoing and shows very promising results, too. In the future it also seems promising to use our algorithm for bootstrapping more sophisticated and thus more expensive algorithms for protein structure determination (such as refining the resulting structures in physics-based force fields similar to~\cite{schug2009,dago2012}).
Moreover, further improvements of the resulting structures could be achieved by re-weighting edges by their density; such an approach would consider surfaces more strongly.


\begin{small}
~\\
\textbf{Acknowledgments.}
The work by MW and HM was partially supported by grant ME 3619/3-1 within German Research Foundation (DFG) Priority Programme 1736.
AS acknowledges support by the \textit{Helmholtz Impuls- und Vernetzungsfonds} and a Google Research Award.
HM and AS acknowledge support by KIT's Young Investigator Network YIN.
The authors thank Michael Kovermann (University of Konstanz) for fruitful discussions. 
\end{small}


\bibliography{references,references-Michael-orig}

\begin{thebibliography}{10}

\bibitem{an2003solving}
Le~Thi~Hoai An.
\newblock Solving large scale molecular distance geometry problems by a
  smoothing technique via the gaussian transform and d.c. programming.
\newblock {\em Journal of Global Optimization}, 27(4):375--397, 2003.
\newblock \href {http://dx.doi.org/10.1023/a:1026016804633}
  {\path{doi:10.1023/a:1026016804633}}.

\bibitem{an2003large}
Le~Thi~Hoai An and Pham~Dinh Tao.
\newblock Large-scale molecular optimization from distance matrices by a d.c.
  optimization approach.
\newblock {\em SIAM Journal on Optimization}, 14(1):77--114, jan 2003.
\newblock \href {http://dx.doi.org/10.1137/s1052623498342794}
  {\path{doi:10.1137/s1052623498342794}}.

\bibitem{DBLP:conf/siamcsc/BergaminiWLM16}
Elisabetta Bergamini, Michael Wegner, Dimitar Lukarski, and Henning Meyerhenke.
\newblock Estimating current-flow closeness centrality with a multigrid
  laplacian solver.
\newblock In {\em Proc. 7th {SIAM} Workshop on Combinatorial Scientific
  Computing, {CSC} 2016}, pages 1--12. {SIAM}, 2016.
\newblock URL: \url{http://dx.doi.org/10.1137/1.9781611974690.ch1}, \href
  {http://dx.doi.org/10.1137/1.9781611974690.ch1}
  {\path{doi:10.1137/1.9781611974690.ch1}}.

\bibitem{berman2000}
Helen~M. Berman, John Westbrook, Zukang Feng, Gary Gilliland, Talapady~N. Bhat,
  Helge Weissig, Ilya~N. Shindyalov, and Philip~E. Bourne.
\newblock The protein data bank.
\newblock {\em Nucleic Acids Research}, 28(1):235--242, Jan 2000.
\newblock \href {http://dx.doi.org/10.1093/nar/28.1.235}
  {\path{doi:10.1093/nar/28.1.235}}.

\bibitem{biswas2008distributed}
Pratik Biswas, Kim-Chuan Toh, and Yinyu Ye.
\newblock A distributed {SDP} approach for large-scale noisy anchor-free graph
  realization with applications to molecular conformation.
\newblock {\em SIAM Journal on Scientific Computing}, 30(3):1251--1277, jan
  2008.
\newblock \href {http://dx.doi.org/10.1137/05062754x}
  {\path{doi:10.1137/05062754x}}.

\bibitem{blumenthal1953dg}
Leonard~M. Blumenthal.
\newblock {\em Theory and Applications of Distance Geometry}, volume 347.
\newblock Oxford, 1953.

\bibitem{brandes}
Ulrik Brandes and Christian Pich.
\newblock Eigensolver methods for progressive multidimensional scaling of large
  data.
\newblock In {\em Graph Drawing}, pages 42--53. Springer Science $+$ Business
  Media, 2007.
\newblock \href {http://dx.doi.org/10.1007/978-3-540-70904-6_6}
  {\path{doi:10.1007/978-3-540-70904-6_6}}.

\bibitem{callahan1995}
Paul~B. Callahan and S.~Rao Kosaraju.
\newblock A decomposition of multidimensional point sets with applications to
  k-nearest-neighbors and n-body potential fields.
\newblock {\em Journal of the {ACM}}, 42(1):67--90, jan 1995.
\newblock URL: \url{http://dx.doi.org/10.1145/200836.200853}, \href
  {http://dx.doi.org/10.1145/200836.200853} {\path{doi:10.1145/200836.200853}}.

\bibitem{crippen1988distance}
Gordon~M Crippen, Timothy~F Havel, et~al.
\newblock {\em Distance Geometry and Molecular Conformation}, volume~74.
\newblock Research Studies Press Taunton, UK, 1988.

\bibitem{dago2012}
Angel~E Dago, Alexander Schug, Andrea Procaccini, James~A Hoch, Martin Weigt,
  and Hendrik Szurmant.
\newblock Structural basis of histidine kinase autophosphorylation deduced by
  integrating genomics, molecular dynamics, and mutagenesis.
\newblock {\em Proceedings of the National Academy of Sciences},
  109(26):E1733--E1742, 2012.

\bibitem{leonadis2015}
Eleonora De~Leonardis, Benjamin Lutz, Sebastian Ratz, Simona Cocco, R{\'e}mi
  Monasson, Alexander Schug, and Martin Weigt.
\newblock Direct-coupling analysis of nucleotide coevolution facilitates rna
  secondary and tertiary structure prediction.
\newblock {\em Nucleic acids research}, 43(21):10444--10455, 2015.

\bibitem{dong2002linear}
Qunfeng Dong and Zhijun Wu.
\newblock A linear-time algorithm for solving the molecular distance geometry
  problem with exact inter-atomic distances.
\newblock {\em Journal of Global Optimization}, 22(1/4):365--375, 2002.
\newblock \href {http://dx.doi.org/10.1023/a:1013857218127}
  {\path{doi:10.1023/a:1013857218127}}.

\bibitem{eades1984}
Peter Eades.
\newblock A heuristic for graph drawing.
\newblock {\em Congressus Numerantium}, 42:146--160, 1984.

\bibitem{fang2013using}
Xingyuan Fang and Kim-Chuan Toh.
\newblock Using a distributed {SDP} approach to solve simulated protein
  molecular conformation problems.
\newblock In {\em Distance Geometry}, pages 351--376. Springer Science $+$
  Business Media, nov 2012.
\newblock \href {http://dx.doi.org/10.1007/978-1-4614-5128-0_17}
  {\path{doi:10.1007/978-1-4614-5128-0_17}}.

\bibitem{fruchterman1991}
Thomas M.~J. Fruchterman and Edward~M. Reingold.
\newblock Graph drawing by force-directed placement.
\newblock {\em Software: Practice and Experience}, 21(11):1129--1164, Nov 1991.
\newblock URL: \url{http://dx.doi.org/10.1002/spe.4380211102}, \href
  {http://dx.doi.org/10.1002/spe.4380211102}
  {\path{doi:10.1002/spe.4380211102}}.

\bibitem{gansner2013}
Emden~R Gansner, Yifan Hu, and Stephen North.
\newblock A maxent-stress model for graph layout.
\newblock {\em IEEE Transactions on Visualization and Computer Graphics},
  19(6):927--940, jun 2013.
\newblock \href {http://dx.doi.org/10.1109/tvcg.2012.299}
  {\path{doi:10.1109/tvcg.2012.299}}.

\bibitem{lange2008}
Oliver~F Lange, Nils-Alexander Lakomek, Christophe Far{\`e}s, Gunnar~F
  Schr{\"o}der, Korvin~FA Walter, Stefan Becker, Jens Meiler, Helmut
  Grubm{\"u}ller, Christian Griesinger, and Bert~L De~Groot.
\newblock Recognition dynamics up to microseconds revealed from an rdc-derived
  ubiquitin ensemble in solution.
\newblock {\em science}, 320(5882):1471--1475, 2008.

\bibitem{leung2009sdp}
Ngai-Hang~Z. Leung and Kim-Chuan Toh.
\newblock An {SDP}-based divide-and-conquer algorithm for large-scale noisy
  anchor-free graph realization.
\newblock {\em SIAM Journal on Scientific Computing}, 31(6):4351--4372, jan
  2010.
\newblock \href {http://dx.doi.org/10.1137/080733103}
  {\path{doi:10.1137/080733103}}.

\bibitem{liberti2009double}
Leo Liberti, Carlile Lavor, Nelson Maculan, and Fabrizio Marinelli.
\newblock Double variable neighbourhood search with smoothing for the molecular
  distance geometry problem.
\newblock {\em Journal of Global Optimization}, 43(2-3):207--218, aug 2009.
\newblock \href {http://dx.doi.org/10.1007/s10898-007-9218-1}
  {\path{doi:10.1007/s10898-007-9218-1}}.

\bibitem{liberti2014survey}
Leo Liberti, Carlile Lavor, Nelson Maculan, and Antonio Mucherino.
\newblock Euclidean distance geometry and applications.
\newblock {\em {SIAM} Review}, 56(1):3--69, jan 2014.
\newblock \href {http://dx.doi.org/10.1137/120875909}
  {\path{doi:10.1137/120875909}}.

\bibitem{lipp2015}
Fabian Lipp, Alexander Wolff, and Johannes Zink.
\newblock Faster force-directed graph drawing with the well-separated pair
  decomposition.
\newblock In {\em Graph Drawing and Network Visualization - 23rd International
  Symposium, {GD} 2015, Los Angeles, CA, USA, September 24-26, 2015, Revised
  Selected Papers}, volume 9411 of {\em LNCS}, pages 52--59. Springer, 2015.
\newblock URL: \url{http://dx.doi.org/10.1007/978-3-319-27261-0_5}, \href
  {http://dx.doi.org/10.1007/978-3-319-27261-0_5}
  {\path{doi:10.1007/978-3-319-27261-0_5}}.

\bibitem{livne2012}
Oren~E. Livne and Achi Brandt.
\newblock Lean algebraic multigrid ({LAMG}): Fast graph laplacian linear
  solver.
\newblock {\em {SIAM} Journal on Scientific Computing}, 34(4):B499--B522, Jan
  2012.
\newblock URL: \url{http://dx.doi.org/10.1137/110843563}, \href
  {http://dx.doi.org/10.1137/110843563} {\path{doi:10.1137/110843563}}.

\bibitem{martin2011}
Jeffrey~W Martin, Anthony~K Yan, Chris Bailey-Kellogg, Pei Zhou, and Bruce~R
  Donald.
\newblock A geometric arrangement algorithm for structure determination of
  symmetric protein homo-oligomers from noes and rdcs.
\newblock {\em Journal of Computational Biology}, 18(11):1507--1523, 2011.

\bibitem{meyerhenke2015}
Henning Meyerhenke, Martin Nöllenburg, and Christian Schulz.
\newblock Drawing large graphs by multilevel maxent-stress optimization.
\newblock In {\em Lecture Notes in Computer Science}, pages 30--43. Springer
  Science $+$ Business Media, 2015.
\newblock URL: \url{http://dx.doi.org/10.1007/978-3-319-27261-0_3}, \href
  {http://dx.doi.org/10.1007/978-3-319-27261-0_3}
  {\path{doi:10.1007/978-3-319-27261-0_3}}.

\bibitem{more1997}
Jorge~J. Mor{\'{e}} and Zhijun Wu.
\newblock Global continuation for distance geometry problems.
\newblock {\em SIAM Journal on Optimization}, 7(3):814--836, aug 1997.
\newblock \href {http://dx.doi.org/10.1137/s1052623495283024}
  {\path{doi:10.1137/s1052623495283024}}.

\bibitem{more1999}
Jorge~J Mor{\'e} and Zhijun Wu.
\newblock Distance geometry optimization for protein structures.
\newblock {\em Journal of Global Optimization}, 15(3):219--234, 1999.

\bibitem{mucherino2012distance}
Antonio Mucherino, Carlile Lavor, Leo Liberti, and Nelson Maculan.
\newblock {\em Distance geometry: theory, methods, and applications}.
\newblock Springer Science \& Business Media, 2012.

\bibitem{noel2013}
Jeffre~K. Noel, Paul~C. Whitford, and Onuchic~Jose N.
\newblock The shadow map: A general contact definition for capturing the
  dynamics of biomolecular folding and function.
\newblock {\em J Phys Chem B}, 116(29):8692--8702, 2013.
\newblock \href {http://dx.doi.org/10.1021/jp300852d}
  {\path{doi:10.1021/jp300852d}}.

\bibitem{ovchinnikov2017}
Sergey Ovchinnikov, Hahnbeom Park, Neha Varghese, Po-Ssu Huang, Georgios~A
  Pavlopoulos, David~E Kim, Hetunandan Kamisetty, Nikos~C Kyrpides, and David
  Baker.
\newblock Protein structure determination using metagenome sequence data.
\newblock {\em Science}, 355(6322):294--298, 2017.

\bibitem{rohl2004}
Carol~A Rohl, Charlie~EM Strauss, Kira~MS Misura, and David Baker.
\newblock Protein structure prediction using rosetta.
\newblock {\em Methods in enzymology}, 383:66--93, 2004.

\bibitem{PDB2015}
Peter~W. Rose, Andreas Prlić, Chunxiao Bi, Wolfgang~F. Bluhm, Cole~H.
  Christie, Shuchismita Dutta, Rachel~Kramer Green, David~S. Goodsell, John~D.
  Westbrook, Jesse Woo, Jasmine Young, Christine Zardecki, Helen~M. Berman,
  Philip~E. Bourne, and Stephen~K. Burley.
\newblock The rcsb protein data bank: views of structural biology for basic and
  applied research and education.
\newblock {\em Nucleic Acids Research}, 43(D1):D345, 2015.
\newblock URL: \url{+ http://dx.doi.org/10.1093/nar/gku1214}, \href
  {http://arxiv.org/abs//oup/backfile/Content_public/Journal/nar/43/D1/10.1093_nar_gku1214/2/gku1214.pdf}
  {\path{arXiv:/oup/backfile/Content_public/Journal/nar/43/D1/10.1093_nar_gku1214/2/gku1214.pdf}},
  \href {http://dx.doi.org/10.1093/nar/gku1214}
  {\path{doi:10.1093/nar/gku1214}}.

\bibitem{saxe1980}
James~B Saxe.
\newblock {\em Embeddability of weighted graphs in k-space is strongly
  NP-hard}.
\newblock Carnegie-Mellon University, Department of Computer Science, 1980.

\bibitem{schug2003}
A~Schug, T~Herges, and W~Wenzel.
\newblock Reproducible protein folding with the stochastic tunneling method.
\newblock {\em Physical review letters}, 91(15):158102, 2003.

\bibitem{schug2010}
Alexander Schug and Jos{\'e}~N Onuchic.
\newblock From protein folding to protein function and biomolecular binding by
  energy landscape theory.
\newblock {\em Current opinion in pharmacology}, 10(6):709--714, 2010.

\bibitem{schug2009}
Alexander Schug, Martin Weigt, Jos{\'e}~N Onuchic, Terence Hwa, and Hendrik
  Szurmant.
\newblock High-resolution protein complexes from integrating genomic
  information with molecular simulation.
\newblock {\em Proceedings of the National Academy of Sciences},
  106(52):22124--22129, 2009.

\bibitem{staudtSM14}
Christian~L. Staudt, Aleksejs Sazonovs, and Henning Meyerhenke.
\newblock Networkit: A tool suite for large-scale complex network analysis.
\newblock {\em Network Science}, 4(4):508--530, Dec 2016.

\bibitem{stothers2012carbon}
JB~Stothers.
\newblock {\em Carbon-13 NMR Spectroscopy: Organic Chemistry, A Series of
  Monographs}, volume~24.
\newblock Elsevier, 2012.

\bibitem{Talbi:2009:MDI:1718024}
El-Ghazali Talbi.
\newblock {\em Metaheuristics: From Design to Implementation}.
\newblock Wiley Publishing, 2009.

\bibitem{uguzzoni2017}
Guido Uguzzoni, Shalini~John Lovis, Francesco Oteri, Alexander Schug, Hendrik
  Szurmant, and Martin Weigt.
\newblock Large-scale identification of coevolution signals across
  homo-oligomeric protein interfaces by direct coupling analysis.
\newblock {\em Proceedings of the National Academy of Sciences},
  114(13):E2662--E2671, 2017.

\bibitem{voet2010biochemistry}
D.~Voet and J.G. Voet.
\newblock {\em Biochemistry, 4th Edition}.
\newblock John Wiley \& Sons, 2010.
\newblock URL: \url{https://books.google.de/books?id=ne0bAAAAQBAJ}.

\bibitem{wuthrich1990}
Kurt W{\"u}thrich.
\newblock Protein structure determination in solution by nmr spectroscopy.
\newblock {\em Journal of Biological Chemistry}, 265(36):22059--22062, 1990.

\end{thebibliography}

\appendix

\section{Appendix}

\subsection{Pseudocode of Simulated Annealing Algorithm}
\label{sub:sa-alg}
%
\begin{algorithm}[!h]
	\KwIn{Graph $G = (V, E),$ distance intervals $d = [l, u]$, $k$, and embedding $x : V \rightarrow \mathbb{R}^k$}
	\KwOut{Embedding $x' : V \rightarrow \mathbb{R}^k$}
	\DontPrintSemicolon
	\SetKwData{T}{t}
	\SetKwData{Change}{change}
	\SetKwData{LocalError}{localError}
	\SetKwData{NewLocalError}{newLocalError}
	\SetKwData{Iterations}{\#iterations}
	\SetKwData{Modifications}{\#modifications}
	
	\SetKwFunction{ComputeLocalError}{computeLocalError}
	\SetKwFunction{LocalForceOptimization}{localForceOptimization}
	\SetKwFunction{Accept}{accept}
	\BlankLine
	
	$x' \leftarrow x$\; 
	\T $\leftarrow 0.3$ \tcp*{Initial temperature}
	\Repeat{\#steps with no improvement $> m$ or \T $< 1e^{-7}$} {
		\While{\Iterations $< 2m$ and \Modifications $< 0.5m$}{
			\ForEach{edge $\{v,w\} \in E$ \emph{\textbf{in parallel}}} {
				\LocalError $\leftarrow$ \ComputeLocalError{$\{v, w\}$, $x'$}\; \label{alg_line:SA_localError}
				$x'_v, x'_w \leftarrow$ \LocalForceOptimization{$v$, $w$, $x'$}\;
				\NewLocalError $\leftarrow$ \ComputeLocalError{$\{v, w\}$, $x'$}\; \label{alg_line:SA_newLocalError}
				\If {not \Accept{\T, \LocalError, \NewLocalError}} { \label{alg_line:SA_accept}
					revert changes made to $x'_v, x'_w$\;
				}
			}		
		}
		
		\T $\leftarrow 0.1 \; \cdot$ \T \tcp*{Cooling}
	} 
	\Return $x'$
	\caption{Simulated annealing algorithm for iDGP} \label{alg:sa_optimizer}
\end{algorithm}


\subsection{Pseudocode of Simple Local Optimizer}
\label{sub:simple-local}
%
\begin{algorithm}[!h]
	\KwIn{Graph $G = (V, E)$, distance intervals $d = [l, u]$, $k$, and embedding $x : V \rightarrow \mathbb{R}^k$}
	\KwOut{Embedding $x' : V \rightarrow \mathbb{R}^k$}
 	\SetKwData{LocalError}{localError}
 	\SetKwData{NewLocalError}{newLocalError}
 	
 	\SetKwFunction{AdjustLength}{adjustLength}
 	\SetKwFunction{ComputeLocalError}{computeLocalError}

	\BlankLine
	
	$x' \leftarrow x$\;
	\For{$i \leftarrow 1$ \KwTo $50$}{		
		remove all locks\; 
		$E' \leftarrow \{ \{v,w\} \in E : \text{error}_{vw}(x')> 10^{-9} \}$\; 
		
		\For{$\{v, w\} \in E'$ in descending order}{
			\If{$\{v, w\}$ is locked} {
				\textbf{continue}\;
			}
			\LocalError $\leftarrow$ \ComputeLocalError{$\{v,w\}$, $x'$}\;
			$x'_v, x'_w \leftarrow$ \AdjustLength{$\{v,w\}$, $x'$}\;
			\NewLocalError $\leftarrow$ \ComputeLocalError{$\{v,w\}$, $x'$}\;
			\If{\NewLocalError $ \geq $ \LocalError}{ 
				revert changes to $x'_v, x'_w$\;
			} \Else {
				lock incident edges of $v$ and $w$
			}			
		}
		
		\If{no improvement} {
			\textbf{break}\;
		}
	}
	\Return $x'$\;	
	\caption{Simple local optimization algorithm.} \label{alg:simple-local}
\end{algorithm}

\newpage

\subsection{Additional Experimental Results for iMDGP}
\label{sub:add-exp}
\begin{table}[!h]
	\caption[Performance results on 70\%-0.1 \mbox{\emph{normal}-iDGP}]{Performance results on  $70\%$ $\sigma=0.1$ \mbox{\emph{normal}-iDGP} instances. 
        Best results in bold font.}    
    \label{tab:t70_0.1}
	\centering
	\nprounddigits{2}
	\begin{tabular}{l | r r r | r r r}
		\hline
		 & \multicolumn{3}{c}{\textbf{\rmsd \ / \AA}} & \multicolumn{3}{c}{\textbf{time \ / s}} \\
		 Protein & \NWK & \DGSOL & \DISCO & \NWK & \DGSOL & \DISCO \\
		\hline
		1ptq & \textbf{\numprint{0.333325215}} & \numprint{7.822672414} & \numprint{0.342835272} & \textbf{\numprint{1.653333333}} & \numprint{9.584444444} & \numprint{17.830} \\
		1bpm & \textbf{\numprint{0.279850229}} & \numprint{20.5765372} & \numprint{0.389025342} &\textbf{ \numprint{12.35222222}} & \numprint{252.0733333} & \numprint{366.400} \\
		1lfb & \textbf{\numprint{0.64064759}} & \numprint{10.57865554} & \numprint{0.927557636} & \textbf{\numprint{2.826666668} }& \numprint{24.25222222} & \numprint{37.97333333} \\
		1toa & \textbf{\numprint{0.288892515}} & \numprint{22.64799532} & \numprint{0.425000273} &\textbf{ \numprint{16.12111111}} & \numprint{237.7644444} & \numprint{394.4466667} \\
		1gpv &\textbf{ \numprint{0.426538405}} & \numprint{14.01915572} & \numprint{1.173984855} & \textbf{\numprint{7.833333333}} & \numprint{116.6844444} & \numprint{181.6988889} \\
		1rgs &\textbf{ \numprint{0.409770431}} & \numprint{16.56987943} & \numprint{0.977945259} &\textbf{ \numprint{7.754444444}} & \numprint{119.4144444} & \numprint{165.0955556} \\
		1f39 & \textbf{\numprint{0.338551401}} & \numprint{16.43962353} & \numprint{0.34786631} & \textbf{\numprint{6.235555556}} & \numprint{77.39333333} & \numprint{84.90444445} \\
		1ax8 & \textbf{\numprint{0.40596479}} & \numprint{11.69402838} & \numprint{0.589752172} & \textbf{\numprint{4.098888889}} & \numprint{40.48777778} & \numprint{66.45666667} \\
		1kdh & \textbf{\numprint{0.755059845} }& \numprint{18.59816985} & \numprint{3.200429192} &\textbf{ \numprint{10.23222222}} & \numprint{185.400} & \numprint{291.2022222} \\
		1mqq & \textbf{\numprint{0.231067377}} & \numprint{23.55308622} & \numprint{0.376605631} &\textbf{ \numprint{19.95222222}} & \numprint{432.7244444} & \numprint{704.6266667} \\
		\hline
	\end{tabular}
\end{table}

\begin{table}[!h]
	\caption[Performance results on 70\%-0.01 \mbox{\emph{normal}-iDGP}]{Performance results on  $70\%$ $\sigma=0.01$  \mbox{\emph{normal}-iDGP} instances.
            Best results (before rounding) in bold font.}
   \label{tab:t70_0.01}
	\centering
	\nprounddigits{2}
	\begin{tabular}{l | r r r | r r r}
		\hline
		 & \multicolumn{3}{c}{\textbf{\rmsd \ / \AA}} & \multicolumn{3}{c}{\textbf{time \ / s}} \\
		Protein & \NWK & \DGSOL & \DISCO & \NWK & \DGSOL & \DISCO \\
		\hline
		1ptq & \numprint{0.27139494} & \numprint{7.025894061} & \textbf{\numprint{0.229313817} }&\textbf{ \numprint{1.568888888} }& \numprint{9.323333333} & \numprint{18.99444444} \\
		1bpm &\textbf{ \numprint{0.215800176}} & \numprint{19.84638816} & \numprint{0.261144947} &\textbf{ \numprint{13.64777778}} & \numprint{196.0522222} & \numprint{385.9677778} \\
		1lfb & \numprint{0.733501796} & \numprint{9.445033908} &\textbf{ \numprint{0.521967753}} &\textbf{ \numprint{3.272222222} }& \numprint{19.220} & \numprint{38.10555556} \\
		1toa & \numprint{0.234706179} & \numprint{22.36035415} & \textbf{\numprint{0.198738045} }&\textbf{ \numprint{16.18777778}} & \numprint{333.2511111} & \numprint{428.730} \\
		1gpv &\textbf{ \numprint{0.370995199}} & \numprint{13.30997954} & \numprint{0.866271987} &\textbf{ \numprint{9.033333333}} & \numprint{102.0066667} & \numprint{192.470} \\
		1rgs & \numprint{0.347975697} & \numprint{16.25028423} &\textbf{ \numprint{0.274118877} }& \textbf{\numprint{8.882222221}} & \numprint{94.50222222} & \numprint{180.220} \\
		1f39 & \numprint{0.260758166} & \numprint{15.30897736} &\textbf{ \numprint{0.184381284}} & \textbf{\numprint{6.273333333}} & \numprint{59.230} & \numprint{94.12888889} \\
		1ax8 & \numprint{0.238135956} & \numprint{10.82422877} & \textbf{\numprint{0.124631977} }&\textbf{ \numprint{4.301111111}} & \numprint{30.82222222} & \numprint{64.98222222} \\
		1kdh & \numprint{0.557582516} & \numprint{19.15400539} &\textbf{ \numprint{0.396341113}} &\textbf{ \numprint{11.18666667} }& \numprint{141.2588889} & \numprint{275.1988889} \\
		1mqq & \textbf{\numprint{0.110699574}} & \numprint{23.48731509} & \numprint{0.111959625} &\textbf{ \numprint{21.060} }& \numprint{506.4122222} & \numprint{827.5522222} \\
		\hline
	\end{tabular}
\end{table}

\begin{table}[!h]
	\caption[Performance results on 70\%-0.001 \mbox{\emph{normal}-iDGP}]{Performance results on  $70\%$ $\sigma=0.001$ \mbox{\emph{normal}-iDGP} instances. 
            Best results (before rounding) in bold font.}
   \label{tab:t70_0.001}
	\centering
	\nprounddigits{2}
	\begin{tabular}{l | r r r | r r r}
		\hline
		 & \multicolumn{3}{c}{\textbf{\rmsd \ / \AA}} & \multicolumn{3}{c}{\textbf{time \ / s}} \\
		 Protein & \NWK & \DGSOL & \DISCO & \NWK & \DGSOL & \DISCO \\
		\hline
		1ptq & \numprint{0.257078061} & \numprint{7.297537935} & \textbf{\numprint{0.196276137} }& \textbf{\numprint{1.834}} & \numprint{6.944444444} & \numprint{19.88777778} \\
		1bpm &\textbf{ \numprint{0.147028981} }& \numprint{19.95338277} & \numprint{0.155261747} & \textbf{\numprint{17.24777778}} & \numprint{158.4088889} & \numprint{386.9011111} \\
		1lfb & \numprint{0.614665085} & \numprint{9.929424217} & \textbf{\numprint{0.547916783}} & \textbf{\numprint{3.681111111}} & \numprint{15.11333333} & \numprint{41.50333333} \\
		1toa &\textbf{ \numprint{0.235413316}} & \numprint{22.08384699} & \numprint{0.238312909} &\textbf{ \numprint{21.80333333}} & \numprint{28.80011111} & \numprint{422.6044444} \\
		1gpv &\textbf{ \numprint{0.385769409} }& \numprint{13.12942481} & \numprint{3.283202799} & \textbf{\numprint{9.616666666}} & \numprint{93.48888889} & \numprint{191.8133333} \\
		1rgs & \textbf{\numprint{0.320799726}} & \numprint{16.94475987} & \numprint{0.332200508} &\textbf{ \numprint{10.490} }& \numprint{86.210} & \numprint{192.910} \\
		1f39 &\textbf{ \numprint{0.186159124} }& \numprint{14.61473417} & \numprint{0.22432632} & \textbf{\numprint{7.915555556} }& \numprint{50.21777778} & \numprint{100.8266667} \\
		1ax8 & \textbf{\numprint{0.243706734}} & \numprint{11.44631047} & \numprint{0.265213958} &\textbf{ \numprint{4.968888888} }& \numprint{26.590} & \numprint{70.63666667} \\
		1kdh & \numprint{0.570181568} & \numprint{17.31079719} & \textbf{\numprint{0.336171664} }& \textbf{\numprint{11.56111111}} & \numprint{120.1466667} & \numprint{316.2377778} \\
		1mqq &\textbf{ \numprint{0.141782809}} & \numprint{23.8222184} & \numprint{0.186888841} &\textbf{ \numprint{27.260}} & \numprint{475.6766667} & \numprint{756.630} \\
		\hline
	\end{tabular}
\end{table}
   
\clearpage 
   
\begin{figure}[h!]
      \centering
       \includegraphics[trim={160px 0 0 160px},width=0.55\textwidth,angle=90]{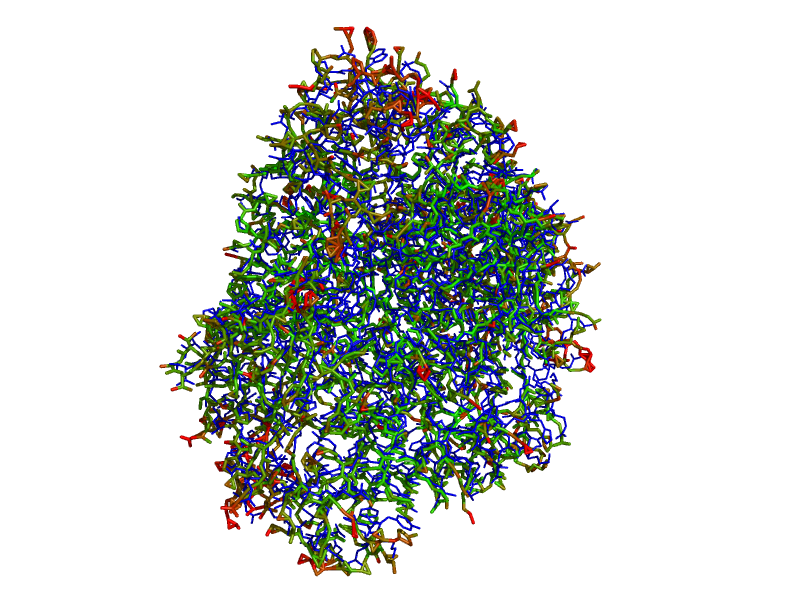}
 
   \caption{Exemplary structure of 1mqq with the reference structure (blue) overlayed with a structure from \NWK for the instance bonds-iDGP (15\% of contact distances, 0.01 noise). Lines are covalent bonds between atoms. Parts that deviate $\geq 4\AA$ are highlighted in red, while parts that agree well are green. The overall \rmsd is 1.42 \AA.
}
   \label{1mqq}
\end{figure}

\begin{figure}[h!]
      \centering
       \includegraphics[trim={160px 0 0 160px},width=0.55\textwidth,angle=90]{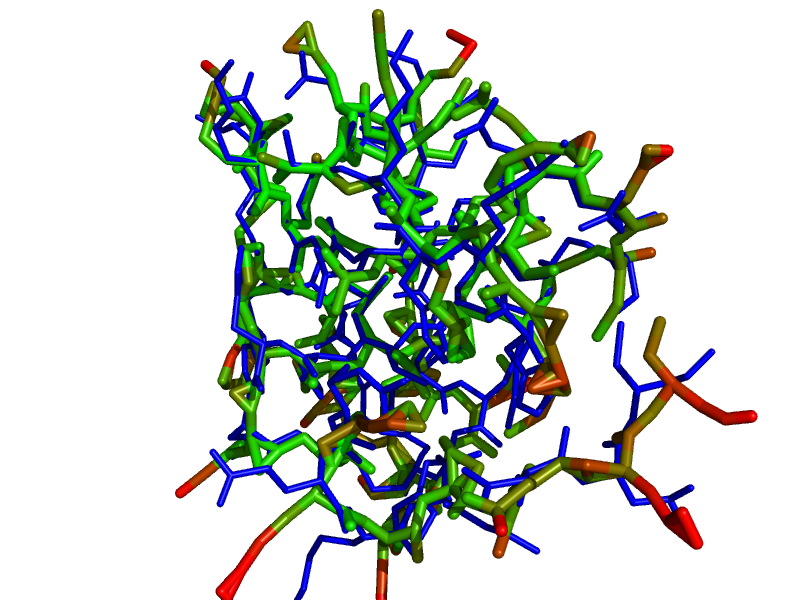}
 
   \caption{Exemplary structure of 1ptq with the reference structure (blue) overlayed with a structure from \NWK for the instance bonds-iDGP (15\% of contact distances, 0.01 noise). Lines are covalent bonds between atoms. Parts that deviate $\geq 4\AA$ are highlighted in red, while parts that agree well are green. The overall \rmsd is 0.99 \AA.
}
   \label{1ptq}
\end{figure}

\clearpage

\subsection{Experimental Results for wiMDGP with \NWK}
\label{sec:add-exp-wiMDGP}
In order to construct the instances, we perform the same procedure as for \mbox{\emph{bonds}-iMDGP} with $50\%$ contact edges. For these contact edges, however, weights and interval bounds are modified into three groups: $25\%$ randomly chosen are considered certain and assigned interval bound of $\pm 0.1$ \AA\ and $c_{vw}=1$, $50\%$  randomly chosen are considered of intermediate confidence and assigned interval bounds of $\sigma=0.1$  and $c_{vw}=0.75$, and the last $25\%$  randomly chosen are considered uncertain and assigned $\sigma=0.5$  and $c_{vw}=0.5$. 

\begin{table}[h!]
\caption{Performance results on wiMPGP instance (bonds + re-weighted $50\%$ contact-edges: $25\%$ $c_{vw}=1$, $\Delta=0.1$\AA\ , $50\%$ $c_{vw}=0.75$, $\sigma=0.1$ ,$25\%$ $c_{vw}=0.5$, $\sigma=0.5$)   }
\label{tab:wiMDGP-t50}
  \centering
	\nprounddigits{2}
	\begin{tabular}{l | r |r }
    	Protein	 &  \rmsd \ / \AA  & time \ / s \\ \hline
1ptq & \numprint{0.604215424112} & \numprint{1.20111111096} \\
1lfb & \numprint{1.0544719929} & \numprint{1.73666666655} \\
1ax8 & \numprint{0.851760041817} & \numprint{2.57111111118} \\
1f39 & \numprint{0.722044262193} & \numprint{1.97000000088} \\
1gpv & \numprint{2.73380542976} & \numprint{1.93444444446} \\
1rgs & \numprint{0.867471333854} & \numprint{5.66333333362} \\
1kdh & \numprint{0.977406999597} & \numprint{7.81000000031} \\
1bpm & \numprint{0.579481085945} & \numprint{11.9066666667} \\
1toa & \numprint{0.600946848525} & \numprint{5.97000000046} \\
1mqq & \numprint{0.497959791515} & \numprint{14.7144444444} \\
	\end{tabular}
\end{table}

\end{document}